\documentclass{caosp309}

\usepackage{graphicx}

\usepackage{natbib}
\articleNo{NNN}
\pubyear{2024}
\volume{NN}
\volnumber{NN}
\firstpage{NN}
\received{MM, DD, YYYY}
\accepted{MM, DD, YYYY}

\def\BibTeX{{\rm B\kern-.05em{\sc i\kern-.025em b}\kern-.08em T\kern-.1667em\lower.7ex\hbox{E}\kern-.125emX}}

\begin{document}

%
\hauthor{J.\,Southworth}

\title{SWIPE: Stars WIth Pulsations and Eclipses}


%
%
\author{J.\,Southworth\inst{1}\orcid{0000-0002-3807-3198}}

%
\institute{Astrophysics Group, Keele University, Staffordshire, ST5 5BG, UK \newline \email{taylorsouthworth@gmail.com}}

\date{Month Day, Year}

\maketitle

\begin{abstract}
Eclipses and pulsations are the two primary ways in which the physical properties of stars can be deduced and used to improve our understanding of stellar theory. An obvious idea is to combine these two analyses into the study of pulsating stars in eclipsing binaries. This is the aim of the SWIPE project. This paper summarises the scientific arguments, current status and future plans of the project.
\end{abstract}

%

\section{Introduction}

Detached eclipsing binary stars (dEBs) are our primary source of direct measurements of the physical properties of stars. It is possible to determine their \emph{bulk properties} -- mass, radius, density, surface gravity -- directly from empirical measurements \citep{Andersen91aarv,Torres++10aarv}. These quantities can be determined to accuracies surpassing 0.2\%, for example AI\,Phe \citep{Maxted+20mn} and V505\,Per \citep{Me21obs5}. They are a crucial check and calibration source for the theoretical stellar models on which modern astrophysics is based. As examples, dEBs have been used to study the strength of internal mixing in high-mass stars \citep{Tkachenko+20aa}, calibrate the mass--radius relation in low-mass stars \citep{Chen+14mn}, help determine the properties of transiting planets \citep{Me09mn}, and as a rung in the cosmological distance ladder \citep{Pietrzynski+19nat,Freedman+20apj}. Unfortunately, very high-precision masses and radii are needed for useful constraints on stellar theory \citep{Valle+16aa,ConstantinoBaraffe18aa}.

Asteroseismology is another prolific route towards understanding the stars, in particular their \emph{interior properties} \citep{Aerts++10book}. Pressure- (p-) mode pulsations are widespread in solar-like oscillators, and $\delta$\,Scuti \citep{Murphy+19mn} and $\beta$\,Cephei \citep{StankovHandler05apjs} stars. Gravity- (g-) mode oscillations manifest in $\gamma$\,Doradus stars \citep{Kaye+99pasp} and the slowly-pulsating B (SPB) stars \citep{Waelkens91aa}. Of these, the g-mode pulsations offer most promise as they penetrate deep inside stars and thus are affected by physical phenomena in the stellar interior. They have been used to constrain the rotational profile of stars \citep{Aerts++19araa}, angular momentum transport \citep[e.g.][]{Mombarg+19mn} and internal mixing \citep{Pedersen+21natas}. A limitation of this method is that the masses and radii of individual stars are not known \emph{a priori} so useful constraints can only be obtained through analysis of significant samples of stars.

An obvious avenue for an improved understanding of the behaviour of stars is to combine the power of binarity and asteroseismology by studying pulsating stars in dEBs. Once masses and radii are precisely known, the measured pulsation frequencies become a sensitive test of theoretical models. Pulsating stars in EBs have been known for decades \citep{Tempesti71ibvs,Knipe71aa}, but the data quality for detailed analysis was only recently achieved with space missions such as \textit{CoRoT, Kepler} and TESS \citep{Me21univ}. Examples of recent work in this area include KIC 9850387 \citep{Sekaran+21aa}, KIC 4142768 \citep{Guo+19apj} and HD 74423 \citep{Handler+20natas}.

We have therefore initiated a project: \emph{Stars WIth Pulsations and Eclipses} (SWIPE) where we aim to perform detailed analysis of a sample of pulsating stars in dEBs to determine their masses, radii, luminosities and oscillation frequencies. These will be used to constrain stellar phenomena such as angular momentum transport, rotational profiles, and internal mixing.

\section{Sample selection}

Early work on pulsating stars in eclipsing binaries (EBs) was hindered by the difficulty of identifying such objects, particularly in cases where the oscillation amplitudes are low and/or the frequencies are comparable to the length of an observing night. The commencement of large-scale ground-based photometric surveys primarily targeted at the detection of transiting extrasolar planets (e.g.\ HATNet, HATSouth, KELT and SuperWASP) has been helpful, but the low data quality and severe aliasing due to diurnal effects compromise the usefulness of observations from these surveys. 

The main advances in this work have instead been realised primarily through space-based data from \textit{CoRoT, Kepler} and TESS. Whilst \textit{Kepler} remains the source of the best light curves, the faintness and relatively low number of its targets hinders work on pulsating dEBs. TESS \citep{Ricker+15jatis} has observed the vast majority of the bright stars in the sky, but in most cases the observations are split into 27-d intervals obtained in different years so have a low frequency resolution. The archives of these space missions are nevertheless suffused with pulsating stars in EBs. \citet{GaulmeGuzik19aa} analysed the full \textit{Kepler} database to find 303 examples, and hundreds of $\delta$\,Scuti stars in EBs have been catalogued \citep{Kahraman+17mn,Kahraman+22raa,Shi+22apjs,Chen+22apjs}.

\citet{MeBowman22mn} presented a catalogue of 26 high-mass pulsators in EBs, based on TESS data and including preliminary analyses of 18 of these objects. This sample of stars was identified visually via inspection of the TESS light curves of known EBs. We found eight objects showing stochastic low-frequency (SLF) variability \citep{Bowman+19natas}, 16 definite or possible $\beta$\,Cephei stars, and 11 definite or possible SPBs. \citet{EzeHandler24apjs} have since presented a catalogue of 78 $\beta$\,Cephei stars in EBs using TESS data.

Whilst there are many $\delta$\,Scuti stars known in EBs (see references above), eclipsing SPB and $\gamma$\,Dor stars appear to be much rarer. This reflects the prevalence of these objects in the single-star population. \citet{MeVanreeth22mn} presented the discovery of $\gamma$\,Dor pulsations in four known dEBs (CM\,Lac, MZ\,Lac, RX\,Dra and V2077\,Cyg), again from visual inspection of TESS data. A fifth, V456\,Cyg, was analysed in more detail as we found it to be only the second object known with tidally perturbed high-order g-mode pulsations \citep{Vanreeth+22aa}. A set of five more EBs showing tidally perturbed g-mode pulsations has been analysed by \citet{Vanreeth+23aa}. We have since discovered more dEBs containing signatures of SLF, $\beta$\,Cephei, SPB, $\delta$\,Scuti and $\gamma$\,Dor pulsations.

\section{A SWIPE status report}

Early results for several objects have been published, concentrating on the higher-mass objects in our sample. Two examples are the first and second $\beta$\,Cephei pulsators with a precise mass measurement: V453\,Cyg and VV\,Ori \citep{Me+20mn,Me++21mn}; in both cases the stars were well-studied dEBs whose TESS data led to the discovery of the pulsational signature. A subsequent study of VV\,Ori by \citet{Budding+24mn} used extensive spectroscopy to revise the mass measurements upwards by a statistically significant amount (masses of $M_1 = 11.56 \pm 0.14$ and $4.81 \pm 0.06$\,M$_\odot$ compared to $M_1 = 10.9 \pm 0.1$ and $4.09 \pm 0.05$\,M$_\odot$ from \citealt{Terrell++07mn}).

An emerging trend is the prevalence of pulsations affected by binarity, specifically showing evidence for tidal excitation, perturbation or tilting. Theoretical analyses of these phenomena have been described by \citet{Fuller+20mn} and \citet{Guo20apj}. Spectroscopic follow-up has been obtained for nine of the objects in the catalogue of \citet{MeBowman22mn} and all five $\gamma$\,Dor EBs named above, among other objects, and analysis of the data for AR\,Cas, AL\,Scl, V453\,Cyg, V446\,Cep and RX\,Dra has begun. Tab.\,\ref{tab:eb} summarises our published work to the present. A basic webpage\footnote{\texttt{https://www.astro.keele.ac.uk/jkt/swipe/}} has been established to document the project, but remains a work in progress. Spectroscopic observations have been obtained for approximately 20 systems in addition to those mentioned above, using the Isaac Newton, William Herschel, Mercator, Nordic Optical and Calar Alto 2.2\,m telescopes 
.

In the near future the PLATO satellite \citep{Rauer+24xxx} will obtain light curves of \textit{Kepler}-like quality over a much larger field of view, to detect Earth-like planets via the transit method. The PLATO Multiple Star Working Group (MSWG) has been established to co-ordinate work on multiple stars using PLATO observations (Southworth \& Maxted, these proceedings).

\begin{table}[t]
\small
\begin{center}
\caption{EBs containing pulsating stars where a reasonably detailed analysis has been published as part of the SWIPE project, given in decreasing order of mass.}
\label{tab:eb}
\begin{tabular}{lllll}
\hline\hline
Target & Pulsation & Spectral type & Period (d) & Ref. \\
\hline                                                                                                    
V1765 Cyg         & SLF                                     & B0.5\,Ib + B1\,V      & 13.374 & (1) \\     
V453 Cyg          & $\beta$\,Cep, tidal                     & B0.4\,IV + B0.7\,IV   &  3.889 & (2) \\     
V346 Cen          & Tidally excited                         & B1\,III + B2\,V       &  6.322 & (3) \\     
VV Ori            & $\beta$\,Cep, tidal						& B1\,V + B4\,V         &  1.485 & (4) \\     
V1388 Ori         & $\beta$\,Cep or SPB                     & B2.5\,IV + B3\,V      &  1.485 & (5) \\     
IQ Per            & SPB                                     & B8\,V + A6\,V         &  1.744 & (6) \\     
KIC 4851217       & $\gamma$\,Dor, $\delta$\,Sct, tidal     & A7\,V + A7\,IV        &  2.470 & (7) \\     
V1229 Tau         & $\delta$\,Sct                           & A0\,Vp(Si) + Am       &  2.461 & (8) \\     
RR Lyn            & $\gamma$\,Dor, $\delta$\,Sct, tidal     & A6\,IVm + F0\,Vm      &  9.945 & (9) \\     
KIC 9851944       & $\gamma$\,Dor, $\delta$\,Sct, tidal     & F1\,V + F2\,IV        &  2.164 & (10) \\    
V456 Cyg          & $\gamma$\,Dor, tidal                    & A2\,Vm + F3\,V        &  0.892 & (11) \\    
GK Dra            & $\gamma$\,Dor, $\delta$\,Sct            & F1\,V + F2\,IV        &  9.974 & (12) \\    
\hline\hline
\end{tabular}
\newline
References: 
(1) \citet{Me23obs6};
(2) \citet{Me+20mn};
(3) \citet{Pavlovski+23aa};
(4) \citet{Me++21mn};
(5) \citet{Me22obs4};
\newline
(6) \citet{Me24obs5};
(7) \citet{Jennings+24mn2};
(8) \citet{Me++23mn};
\newline
(9) \citet{Me21obs6};
(10) \citet{Jennings+24mn};
(11) \citet{Vanreeth+22aa};
\newline
(12) \citet{Me24obs1}.
\end{center}
\end{table}

\acknowledgements
\noindent We acknowledge support from the Science and Technology Facilities Council (STFC) under grant number ST/Y002563/1.

\end{document}